\documentclass[]{spie}  
\usepackage[]{graphicx}
\usepackage[]{amsmath}
\usepackage[]{aas_macros}
\usepackage[T1]{fontenc}

\newcommand{\Spider}{\textsc{Spider}}

\newcommand{\degr}{\hbox{$^\circ$}}

\newcommand{\reg}{\textsuperscript{\textregistered}}
\newcommand{\tm}{\textsuperscript{\texttrademark}}

\title{Design and construction of a carbon fiber gondola for the {\LARGE \textsc{\bfseries Spider}} balloon-borne telescope} 

\author{J.$\,$D. Soler\supit{a,b}\supit{*}, P.~A.~R. Ade\supit{c}, M. Amiri\supit{d}, S.~J. Benton\supit{e}, J.~J. Bock\supit{f}, J.~R. Bond\supit{g,h}, S.~A. Bryan\supit{j}, C. Chiang\supit{k,m}, C.~C. Contaldi\supit{n}, B.~P. Crill\supit{f,i}, O.~P. Dor\'{e}\supit{ f,i}, M. Farhang\supit{g}, J.~P. Filippini\supit{f}, L.~M. Fissel\supit{b}, A.~A. Fraisse\supit{k}, A.~E. Gambrel\supit{k}, N.~N. Gandilo\supit{b}, S. Golwala\supit{f}, J.~E. Gudmundsson\supit{k}, M. Halpern\supit{d,h}, M. Hasselfield\supit{d,k}, G.~C. Hilton\supit{o}, W.~A. Holmes\supit{i}, V.~V. Hristov\supit{f}, K.~D. Irwin\supit{o}, W.~C. Jones\supit{k}, Z.~D. Kermish\supit{k}, C.-L. Kuo\supit{l}, C.~J. MacTavish\supit{g}, P.~V. Mason\supit{f}, K.~G. Megerian\supit{i}, L. Moncelsi\supit{f}, T. Morford\supit{f}, J.~M. Nagy\supit{j}, C.~B. Netterfield\supit{b,e,h}, R. O'Brient\supit{f,i}, A.~S. Rahlin\supit{k}, C.~D. Reintsema\supit{o}, J.~E. Ruhl\supit{j}, M.~C. Runyan\supit{i}, J.~A. Shariff\supit{b}, A. Trangsrud\supit{i}, C. Tucker\supit{c}, R.~S. Tucker\supit{f}, A.~D. Turner\supit{i}, A.~C. Weber\supit{i}, D.~V. Wiebe\supit{d}, E.~Y. Young\supit{k}
\skiplinehalf
\supit{a}Institute d'Astrophysique Spatiale. CNRS. France;\\
\supit{b}Department of Astronomy and Astrophysics. University of Toronto. Canada; \\
\supit{c}School of Physics \& Astronomy. Cardiff University. UK; \\
\supit{d}Department of Physics \& Astronomy. University of British Columbia. Canada; \\
\supit{e}Department of Physics. University of Toronto. Canada; \\
\supit{f}Department of Physics. California Institute of Technology. USA; \\
\supit{g}CITA. University of Toronto. Canada; \\
\supit{h}Canadian Institute for Advanced Research. Canada; \\
\supit{i}Jet Propulsion Laboratory. USA; \\
\supit{j}Department of Physics. Case Western Reserve University. USA; \\
\supit{k}Department of Physics. Princeton University. USA; \\
\supit{l}Department of Physics. Stanford University. USA; \\
\supit{m}School of Mathematics, Statistics \& Computer Science. University of KwaZulu-Natal. South Africa; \\
\supit{n}Theoretical Physics Blackett Laboratory. Imperial College. UK; \\
\supit{o}National Institute of Standards and Technology. USA;
}

\authorinfo{\supit{*}Corresponding author; E-mail: jsolerpu@ias.u-psud.fr, Telephone: +33 1 69 85 86 20\\
Copyright 2014 Society of Photo-Optical Instrumentation Engineers. One print or electronic copy may be made for personal use only. Systematic reproduction and distribution, duplication of any material in this paper for a fee or for commercial purposes, or modification of the content of the paper are prohibited.}
 
\begin{document} 
\maketitle 

\begin{abstract}
We introduce the light-weight carbon fiber and aluminum gondola designed for the \Spider\ balloon-borne telescope. \Spider\ is designed to measure the polarization of the Cosmic Microwave Background radiation with unprecedented sensitivity and control of systematics in search of the imprint of inflation: a period of exponential expansion in the early Universe. The requirements of this balloon-borne instrument put tight constrains on the mass budget of the payload. The \Spider\ gondola is designed to house the experiment and guarantee its operational and structural integrity during its balloon-borne flight, while using less than 10\% of the total mass of the payload. 
We present a construction method for the gondola based on carbon fiber reinforced polymer tubes with aluminum inserts and aluminum multi-tube joints. We describe the validation of the model through Finite Element Analysis and mechanical tests.   
\end{abstract}



\section{INTRODUCTION}\label{sec:intro}

Balloon-borne experiments are limited in mass by the buoyancy of the helium balloons used to lift them. The mass available for scientific equipment on board the payload is constrained by the mass of the structural elements that guarantee the integrity of the experiment during the balloon-borne flight. The structure of an ideal payload must therefore combine durability and high strength per unit of mass. The optimization for high strength and low weight is possible using composite materials and aggressive light-weighting techniques derived from the results of detailed Finite Element Analysis (FEA).

This document describes the design of the \Spider\ gondola: a pointed platform, which combines composite materials with light-weighted aluminum elements to provide support for the \Spider\ instrument \cite{filippini2010,fraisse2013,rahlin2014}. The design of the \Spider\ gondola was motivated by the large volume cryogenic vessel necessary to achieve the goals of the experiment \cite{fraisse2013,odea2011}. The structure was custom made to accommodate the cryostat, which houses a set of six telescopes. This cryostat is larger and heavier than any previous balloon-borne experiment cryostats\cite{crill2003,maciasperez2007,fissel2010,reichborn2010}. The \Spider\ gondola also allows the motion of the cryostat in azimuth and elevation and protects its structural integrity during flight and after termination. Some of the products of the \Spider\ design such as the carbon fiber and aluminum sunshields as well as the pivot have successfully flown in the BLASTPol Long Duration Balloon (LDB) flight from Antarctica in December 2010 (BLASTPol10) and 2012 (BLASTPol12)\cite{fissel2010, pascale2008}.

The technique used for the design of the \Spider\ gondola is modular and can be extended to future balloon-borne experiments. Many of the parts, such as the outer frame multi-tube joints, are custom made but can be easily adapted to other payload geometries. Others, such as the sunshield hubs and inserts, can be used to produce structures with multiple geometries. The motor assemblies and elevation drive sets can be scaled to accommodate the necessities of other balloon-borne instruments. The specific requirements of the gondola change from experiment to experiment and even from campaign to campaign. However, the state of the art Computer-Assisted Design (CAD) tools and novel composite materials used in the \Spider\ design can be applied to a broad range of experiments.

This proceeding is organized as follows: Section \ref{mecha:intro} describes the structural requirements for the balloon-borne platform and presents the critical scenarios considered for its design. Section \ref{mecha:suspensionelements} introduces the suspension elements, which link the payload to the flight train. Section \ref{mecha:outerframe} describes the design and construction of the carbon fiber and aluminum structure composing the \Spider\ outer frame. Finally, Section \ref{mecha:sunshields} presents the design and construction techniques used for the sunshield frames.

\begin{figure}
\centerline{\includegraphics[height=0.25\textheight]{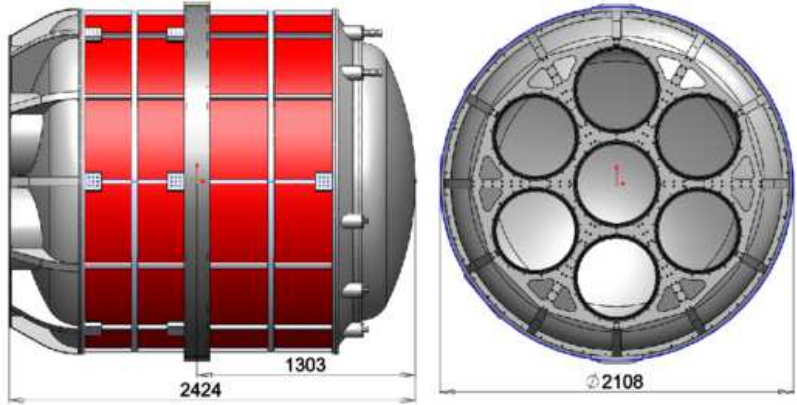}}
\caption{Rendering of the \Spider\ LDB cryostat. The corresponding dimensions are in millimeters.}\label{mecha:spidercryostat}
\end{figure}

\section{Design Benchmarks}\label{mecha:intro}

The size of the \Spider\ gondola is directly related to the scientific goals of the experiment. The size of the \Spider\ refracting telescopes is determined by the angular resolution adequate for the observation of the imprint of primordial gravitational waves in the polarization of the Cosmic Microwave Background (CMB) polarization\cite{seljak1997}. The size of the cryostat is determined by the diameter and length of the telescope inserts plus the volume necessary to hold enough cryogens to achieve the necessary integration times. In sum, the size of the gondola is defined by the dimensions of the cryostat, illustrated in Figure \ref{mecha:spidercryostat}.

\begin{figure}
\centerline{\includegraphics[width=0.85\linewidth]{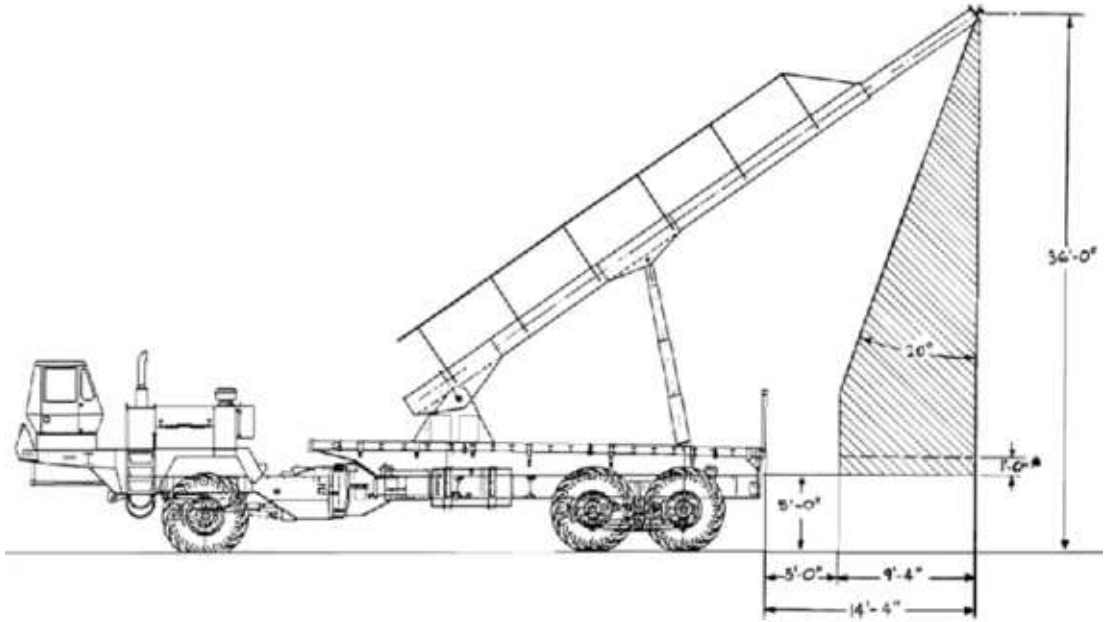}}
\caption[Geometric Limits for Gondola Design]{Geometrical limits for the gondola design set by the Columbia Scientific Balloon Facility (CSBF) deployment vehicle in Antarctica \cite{CSBF2011}.}\label{LaunchVehicle}
\end{figure}

Further geometrical limits are set by the space in the high bay where the gondola is assembled in Antarctica and the limits of the deployment vehicle. The launch procedure requires that no part of the gondola intersects a plane 20\degr\ from the vertical axis passing through the attachment point between the launch vehicle and the gondola as shown in Figure \ref{LaunchVehicle}. Additionally, in order to transport the gondola to the integration and launch location, the separate components of the gondola must fit in standard shipping containers with interior dimensions 2.38~m in height by 2.34$\,$m in width, and can be 5.71$\,$m or 12.03$\,$m deep according to the International Organization for Standardization (ISO) standard 6346.

The balloons and the facilities necessary to fly \Spider\ are provided by the Columbia Scientific Balloon Facility (CSBF), a division of the National Aeronautics and Space Administration (NASA) dedicated to balloon flights. The maximum mass of the scientific payload is approximately 2268$\,$kg (5000$\,$lb), which corresponds to the maximum gross lift of the balloon (8000$\,$lb) minus the mass of the balloon itself, the flight train, the Support Instrumentation Package (SIP), the solar arrays, the ballast hopper and enough ballast to maintain the altitude during the duration of the flight. The dry mass of the \Spider\ cryostat is roughly $\sim$1600$\,$kg (3500$\,$lb) when fully integrated and loaded with cryogens. The remaining 668$\,$kg (1500$\,$lb) are distributed between the motorized systems, which allow pointing in azimuth and elevation, the Attitude Control System (ACS), the flight computers, the serial hub, batteries, the solar array, pointing sensors, sunshields, miscellaneous electronic boxes, and the frame, which supports the whole assembly.

\subsection{The \Spider\ Gondola}
\begin{figure}
\centerline{
\includegraphics[height=0.42\textheight]{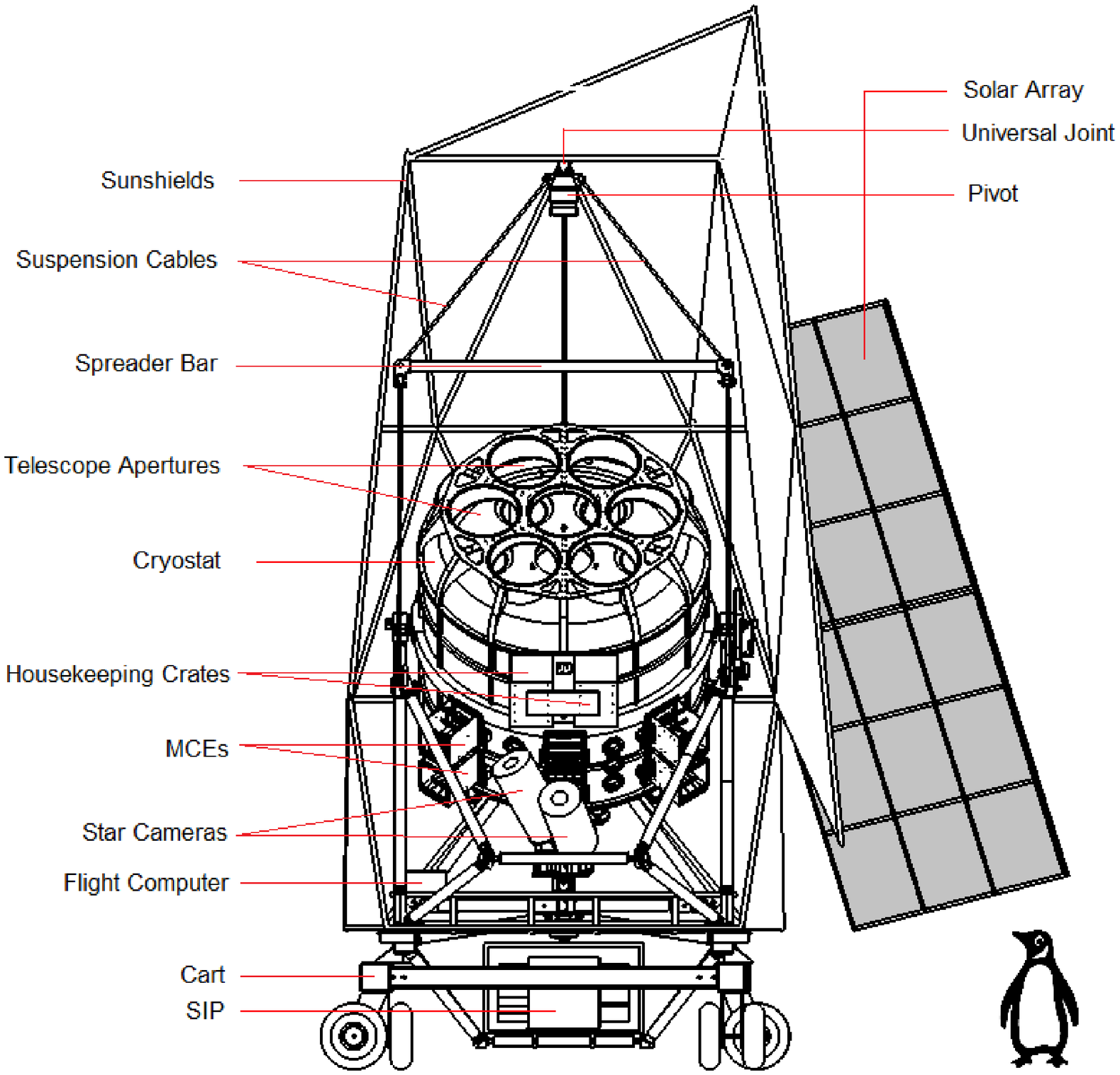}
\includegraphics[height=0.42\textheight]{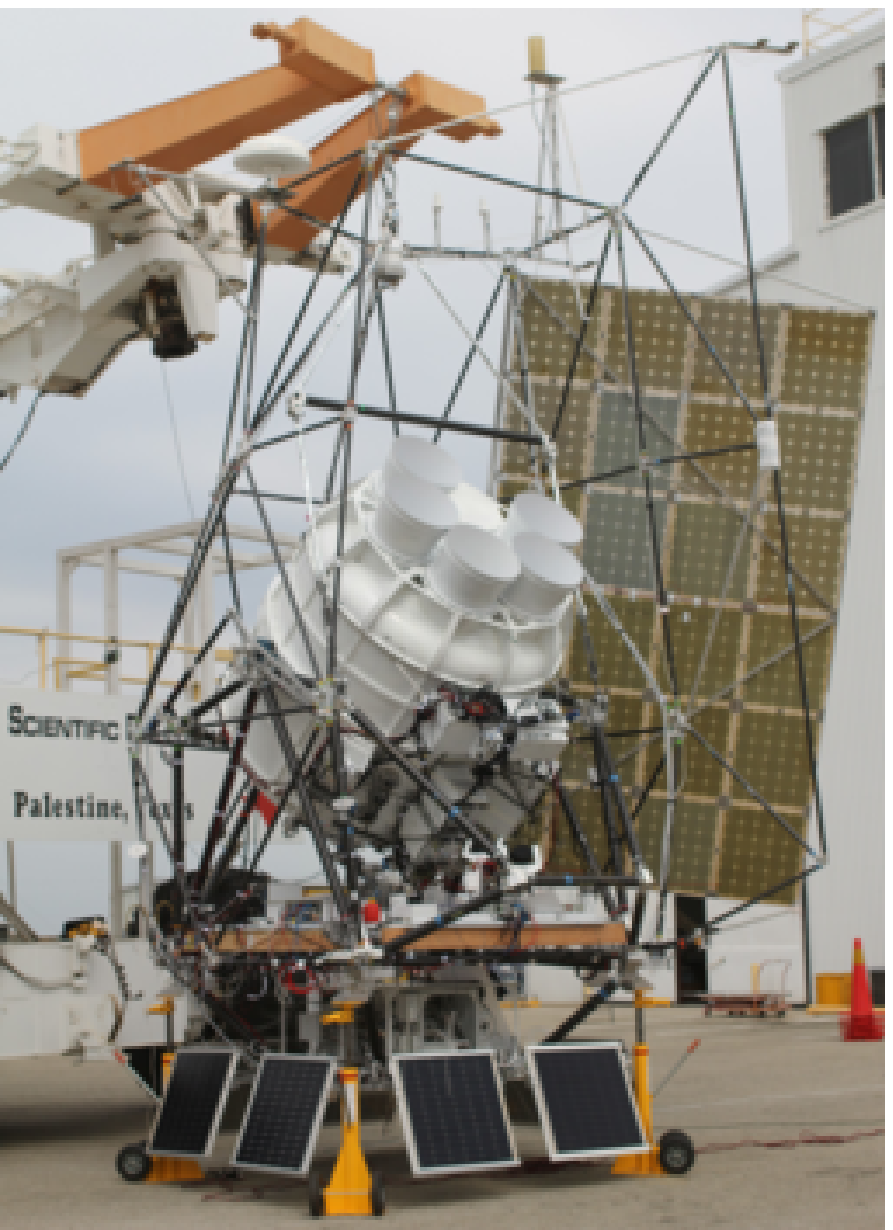}}
\caption{Left: Schematic drawing of the \Spider\ gondola including a 1$\,$m-tall penguin for scale. Right: photograph of the \Spider\ gondola fully assembled during CSBF compatibility test in 2013.}\label{FullGondola}
\end{figure}

The \Spider\ payload, shown fully assembled during compatibility test in Figure \ref{FullGondola}, is composed of three main parts: the outer frame or gondola, made out of aluminum and Carbon Fiber Reinforced Polymer (CFRP) tubes; the cryostat, which is effectively an inner frame trunnion-mounted onto the outer frame at two points along a horizontal axis; and a set of sunshields, which attaches to the outer frame.

The payload is designed to work in two configurations. In the \emph{laboratory configuration}, the gondola sits on top of a customized aluminum cart with pneumatic tires, which allow moving the gondola without the need of a hoist. In the \emph{flight configuration}, the pivot hangs from a hoist and the gondola is suspended from the cables. The blocks attaching the outer frame to the cart are removed and the ballast hooper and the solar arrays are attached to the square frame beneath the SIP.

\begin{figure}
\centerline{\includegraphics[height=0.4\textheight]{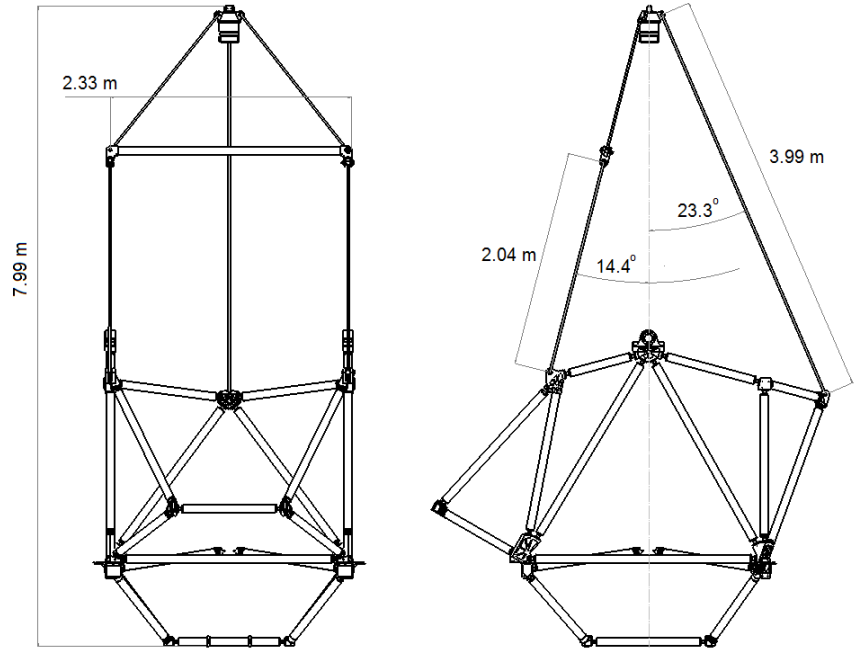}}
\caption{Position of suspension cables and hanging angles indicated on a front (shown in the left hand side) and a right view of the \Spider\ gondola.}\label{mecha:suspensionangles}
\end{figure}

The outer frame is a customized truss structure composed of CFRP tubes with aluminum inserts, which are fastened together by multi-tube aluminum joints. It is suspended from the pivot by three cables as illustrated in Figure \ref{mecha:suspensionangles}. All the beams in the structure are distributed in triangles in order to maintain axial forces and minimize moment on truss elements. All angles and distances have been chosen in order to locate the payload elements and optimize the force distribution. The technique used to construct the outer frame is described in Section \ref{mecha:outerframe}.

The outer frame geometry allows 360\degr\ rotation of the cryostat around the trunnions. This feature facilitates the integration of the gondola and the installation of the telescope inserts in the cryostat. During the flight, the elevation limits are set by the elevation drive. The telescope can point between 20\degr\ and 50\degr\ in elevation. The minimum elevation corresponds to the orientation where the telescope observes the ground. The maximum elevation is the orientation where the telescope sees the balloon. Both values are obtained by modeling the \Spider\ beams as a set of truncated cones with a 10\degr\ aperture and coincident with the telescope apertures.

The entire gondola can rotate to any azimuthal angle. The center-of-mass of the system is on the rotational axis so that translation of the gondola does not generate torques that re-orient the telescope. The limits of observation in azimuth are constrained by the exposure of the telescope to direct sunlight. Therefore, the geometry of the sunshields defines the area of the sky, which can be observed by the experiment \cite{soler2014therm}.

The \Spider\ cryostat was designed and manufactured by RedStone Aerospace. The structural support of the interior components of the cryostat relies on G10/aluminum flexures symmetrically placed on the cylinder sides, as described in previous SPIE proceedings\cite{gudmundsson2010, runyan2010}. The analysis of the structural integrity of the interior of the cryostat is beyond the scope of this document but it is described on the Redstone Aerospace design reports\cite{redstone2008}. 

The \Spider\ sunshield is also a truss structure composed of CFRP tubes with aluminum inserts, which are fastened together by multi-tube aluminum joints. The sunshields have less demanding structural requirements than the outer frame; thus the technique used for their construction is modular and less restrictive as discussed in Section \ref{mecha:sunshields}.

\subsection{Critical Design Scenarios}\label{DesignScenarios}

The geometry of the outer frame was defined through a series of beam mesh simulations made with the SolidWorks\tm\ Simulation package. In these simulations, the frame is modeled by a series of truss elements, which have specified cross sections and material properties. The trusses are connected with ideal nodes, which constrain rotation or translation depending on the design scenario. Gravity and accelerations can be defined in the model and loads can be applied to the nodes and trusses. The result of the beam mesh simulation is a load table, which presents the axial forces, bending forces, and stresses on multiple segments of each truss. The beam mesh model is run and tested in a set of critical design scenarios defined by large accelerations and shocks suffered by the structure during the balloon-borne flight. 

Initially, the beam mesh synthetic model is used for optimizing the distribution of the truss elements. The resulting load tables are used to test and select the materials of the frame. The selection of the CFRP tubes in the \Spider\ outer frame is made using a series of beam mesh simulations using tubes of different diameters and wall thicknesses. Once the tubes are selected, the load tables from each study are used as an input for the simulation of individual joints and inserts. The following are the critical scenarios considered in the design of \Spider.

\subsubsection{Chute Shock}\label{DS:ParachuteShock}

All structural components of the gondola must survive a vertical acceleration of 97.8$\,$m$\,$s$^{-2}$ (10$\,$g). Such acceleration results from the parachute opening and breaking the free-fall of the gondola after balloon termination. The pull produced by the parachute is named chute-shock, and it is described as ``\emph{a load 10 times the weight of the payload applied vertically at the suspension point}'' according to the CSBF structural requirements for balloon gondolas \cite{CSBF2011}. This is the dominant scenario in the design considerations of the \Spider\ gondola as it is the one, which translates into largest forces on the structural elements.

The simulation of this scenario was made using two redundant models. The first was made by applying the forces corresponding to the tension on the suspension cables to the corresponding nodes and constraining the nodes and trusses where the main loads are located. The second was made by assigning the mass of the different components to their corresponding support trusses and nodes and then applying a vertical acceleration of 97.8~m~s$^{-2}$. Both methods overestimate the stresses on the frame, so their results are adequate for validating the design. The results of the first simulation produce slightly larger values of stress in the truss elements and those are the values used for validating the model.

The results of the simulation give a minimum Factor of Safety (FOS) of 4.14 for gondola structural elements, when comparing the tensile strength of the CFRP tubes (1,896.05$\,$MPa) to the maximum stress. This minimum FOS corresponds to the tensile stress on both of the tubes directly below the front suspension cables. 

The effect of the chute shock on the cryostat was included in the design specifications used by RedStone Aerospace for the design of the \Spider\ cryostat. Each trunnion mount has a FOS of 1.24. The minimum FOS in the suspension cables and the spreader bar are 2.85 and 3.79 respectively.

\subsubsection{Pin release (Uneven loading)}

At launch, the gondola will be suspended from a pin located on the end of the arm of the deployment vehicle. The vehicle moves in order to align the payload beneath the balloon before releasing. However, the wind can displace the balloon to form an angle of a few degrees with the vertical axis. In a gondola suspended by cables, such as \Spider, the misalignment results in the momentary concentration of the load in one or one pair of the suspension cables and subsequently an abrupt pull produced by the flight train catching the payload after it is released from the pin. The result of this violent motion is a large angular acceleration of the cryostat. This angular acceleration was estimated to be around 600\degr\ s$^{-2}$ in the launch of the BLAST LDB flight from Antarctica in December 2006 (BLAST06). This sudden rotation damaged the lock pin in the BLAST LDB flight from Kiruna, Sweden in 2005 (BLAST05), and destroyed the elevation drive in the 2009 flight of the E and B Experiment (EBEX)\cite{reichborn2010}.

The prescription given by CSBF for treating this scenario is: ``\emph{a load five times the weight of the payload applied at the suspension point and 45\degr\ to the vertical ... if flexible cable suspension systems are used, they must be able to withstand uneven loading caused by cable buckling}''. The worst scenario in case of cable buckling would be having the whole payload hanging momentarily from a single cable, a situation in which the minimum FOS is 1.10 for the \Spider\ suspension cables.

For the \Spider\ gondola, the effect of a 600\degr\ s$^{-2}$ angular acceleration is mainly received by three elements: the cryostat assembly, the elevation drive, which locks the cryostat in a fixed elevation position, and the outer frame, which ultimately receives the impact of the accelerating cryostat. The cryostat assembly is supported by G10 flexures designed to support the internal components of the cryostat during the chute shock. The flexures located in the extremes of the cylinder are located at 1$\,$m from the rotational axis and have enough mechanical advantage to support the internal components.

The elevation drive-locking pin of \Spider\ is located 1~m from the rotation axis of the cryostat. This distance was chosen to increase the mechanical advantage of the pin and minimize the effect of the shock on the gondola. This locking strategy produces minimum FOS of 5.14, 2.16, and 1.48 for the cryostat, the trunnion and the locking arm respectively \cite{shariff2014}.

The effect of the deployment vehicle pin release on the outer frame is modeled by applying the force of the elevation drive-locking pin on the corresponding node. This force is calculated using the torque resulting from the cryostat moment of inertia and the angular acceleration. The simulation of this load on the outer frame gives a minimum safety factor of 2.45 on the truss elements. Additionally, we calculated the effect of a lateral acceleration of 5g on the gondola elements. The model of the outer frame of the \Spider\ gondola has a minimum safety factor of 4.1 for the truss elements in this case. The FOS considerations on the cryostat are redundant with the results of the 10g analysis.

\subsubsection{Landing: Resting on the base at 5g}\label{DS:Landing}

The surface winds in Antarctica make an upright landing very unlikely, even for a gondola with legs like BLASTPol \cite{fissel2010}. Consequently, to save weight, no attempt was made in the design of \Spider\ to prevent the payload from rolling over onto one of its sides after landing. Mechanically, the most demanding instant during such a landing is the first contact with the ground. In this stage the most exposed elements of the gondola are the lower frame and the supporting elements of the SIP. In order to simulate these landing conditions, the gondola was required to withstand a load equivalent to its mass at 5g while resting on its lower plane. The results of the simulation give a FOS of 3.77 on the truss elements.

\subsubsection{On cart at 1g}\label{DS:OnCart}

During integration, the outer frame is fastened to the cart on the nodes in the corners of the upper square frame. The loads produced by the cryostat, the boxes and the SIP have been tested at 10g on the outer frame in the parachute shock scenario. Given that the support points are different, a new simulation was made to guarantee the structural integrity of the payload during integration. This case is not as critical as the chute shock but it is very important given that scientists and non-flight equipment are going to be around the structure during the integration. With an additional load of 200$\,$kg (441\,lb) on the deck, the FOS of the truss elements is larger than 10.

\subsubsection{Frequency Analysis}

\begin{figure}
\centerline{\includegraphics[height=0.25\textheight]{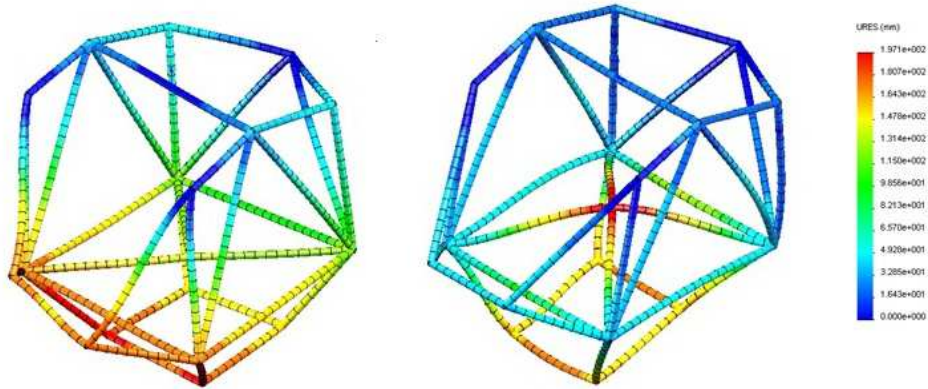}}
  \caption{Displacement diagram corresponding to the first two resonance modes of the \Spider\ outer frame at 29.45 (left) and 81.50$\,$Hz (right).}\label{mecha:resonantmodes}

\end{figure}

As a consequence of the pointing requirements, the minimum feedback rate of the control system is 10$\,$Hz. To accommodate this, the outer frame is designed to be rigid with resonant frequencies over that value. The minimum resonant frequency of the outer frame is located at 29.45$\,$Hz and the next harmonic is at 81.50$\,$Hz. These resonant modes are show in Figure \ref{mecha:resonantmodes} and they are calculated from a model of vibration response made with a static beam mesh model in SolidWorks\tm.

The \Spider\ outer frame profits from the excellent vibration damping provided by the fiber winding in the CFRP. Similar to the behavior of handle structural tubes in mountain and road bikes, the \Spider\ tube mesh absorbs the vibrations of the structure. Nevertheless, particular attention has been devoted to maintaining all mechanical tolerances to minimize backlash from motors and to servicing the bearing units with low-temperature grease to avoid further vibrations.

\section{Suspension elements}\label{mecha:suspensionelements}

The \Spider\ gondola is attached to a steel wire ladder that composes the flight train using a customized universal joint. The universal joint is attached at one end to the ladder and to the pivot shaft at the other end. The shaft rotates inside a steel casing supported by a thrust bearing. Each of the three suspension cables is attached to steel tabs welded to a steel casing. The suspension cables end in thimbles, which are fastened to the tabs by round pin anchor shackles. The two front cables are attached to a carbon fiber and aluminum spreader bar that avoids the effect of lateral forces on the outer frame. A single cable directly attaches the pivot to the back of the gondola as illustrated in Figure \ref{mecha:suspensionangles}.

\subsection{Universal Joint}

\begin{figure}
\centerline{\includegraphics[height=0.3\textheight]{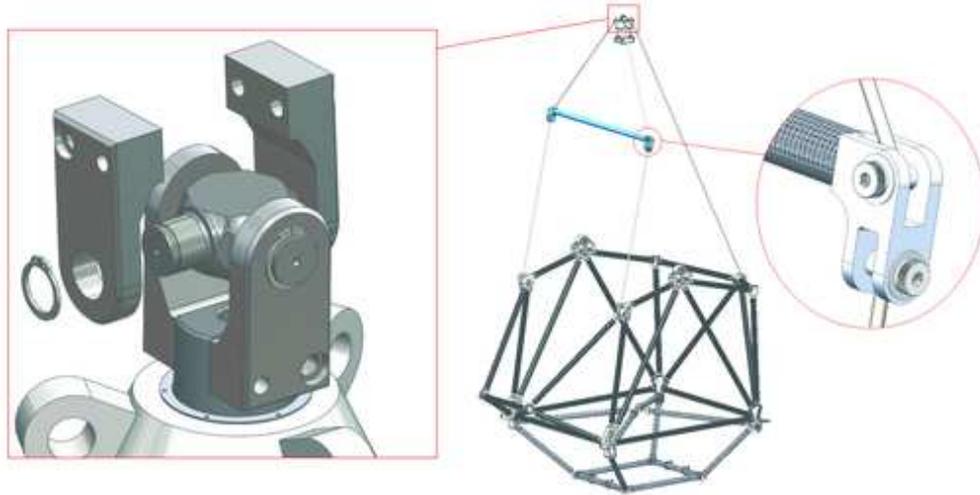}}
\caption{Left: Exploded view of the \Spider\ Universal Joint. Part of the pivot steel casing is visible at the bottom. The upper segment of the universal joint is fastened to a coupling block provide by CSBF. Right: Spreader Bar.}\label{mecha:universaljoint}
\end{figure}

The universal joint used on \Spider, shown in Figure \ref{mecha:universaljoint}, is a reproduction of the design successfully flown in the BLAST test flight from Fort Sumner, NM in 2003 (BLAST03), BLAST05, BLASTPol10, and BLASTPol12 and based on the original design by AMEC Dynamic Structures. It is composed of a monolithic steel core with four cylindrical shafts where two bored aluminum yokes are free to rotate. A steel snap ring at the end of each shaft retains the aluminum yoke. One of the aluminum yokes connect to the flight train while the other connects to the pivot shaft. The two-rotation-axis design of the universal joint allows free rotation up to 135\degr\ in each direction. The larger thermal expansion coefficient of aluminum compared to steel guarantees that the side elements are free to rotate even when exposed to the temperature changes experienced during the LDB flight. An initial design of the \Spider\ universal joint included oil-lubricated phosphor bronze flanged sleeve bearings to reduce the friction between the steel shaft and the aluminum bore. However, given the demanding pointing control requirements, the latest version of the universal joint uses a set of needle bearings\cite{shariff2014}. At chute shock, the steel core and aluminum plates of the universal joint have FOSs of 7.48 and 3.22 respectively.

\subsection{Suspension Cables}

The \Spider\ gondola is suspended by cables made with braided Technora\reg, a high modulus polyamide known for its high strength, heat resistance, low stretch, and better resistance to flex-fatigue than other high modulus aramid fibers \cite{syntheticropes}. The Technora\reg\ cables are considerably lighter than steel ropes of equivalent strength and make integration operations more manageable. The degradation of the synthetic fibers by exposure to ultraviolet (UV) light is prevented using an aluminized mylar sleeve. This sleeve also prevents overheating, which would produce thermal degradation of the fibers.

The \Spider\ suspension ropes are custom made by Helinets, a supplier of helicopter cargo lines. \Spider\ uses five 5/8~inch diameter segments for the back and front suspension cables. Each rope is hand braided and pre-stretched. This process results in length tolerance of around 0.30$\,$m (1$\,$ft). For this reason, the synthetic cables have to be combined with turnbuckles that allow adjusting the pin-to-pin distances.

The 5/8 and 3/4$\,$inch ropes are rated for maximum axial loads of 38,700$\,$lbf and 58,500$\,$lbf. The minimum FOS of the suspension ropes is determined by CSBF: ``\emph{Each cable, cable termination and cable attachment must have an ultimate strength greater than five times the weight of the payload divided by the sine of the angle that the cable makes with horizontal, which should be larger than 30\degr, in a normal flight configuration.}''. The FOS and nominal length of each cable are summarized in Table \ref{mecha:suspensionropesTable}.

\begin{table}[htbp!]
\caption{Minimum safety factors on the \Spider\ gondola suspension cables. The pin-to-pin distance refers to the nominal distance between the coupling points and thus it corresponds to the length of the rope plus the turnbuckle plus the shackle.}\label{mecha:suspensionropesTable}
\begin{center} 
\begin{tabular}{|c|c|c|c|}
    \hline
    Rope            & Pin-to-pin & Load   & FOS \\
                    & length (m)    & at 5g  &           \\ \hline
    Back            & 3.997   & 26,416  & 1.47      \\
Front top       & 1.810   & 31,063  & 1.25      \\    
Front bottom    & 2.094   & 25,511  & 1.50      \\
    \hline
\end{tabular}
\end{center}
\end{table} 

\subsection{Spreader Bar}

A spreader bar cancels the horizontal components of the tension on the suspension cables to minimize the lateral forces acting on the outer frame structure. The \Spider\ spreader bar consists of a CFRP tube and two aluminum inserts that attach to the closed spelter socket end of the suspension cables as shown in Figure \ref{mecha:universaljoint}.

The CFRP tubing was a straightforward solution for the spreader bar given the analysis made for the design of the outer frame, which will be described in Section \ref{mecha:outerframe}. Given the role of the spreader bar, the design goal of the inserts is to maintain the force produced by the suspension cables acting on the axis of the tube, and this was achieved by constructing the part around the node where the axis of the cables and the axis of the tube intersect. The minimum safety factor obtained from simulations of the inserts is 1.28. 

Since the design of the inserts guarantees on-axis force, the main failure mode for the CFRP tube is Euler buckling. Euler buckling, also known as elastic instability, is characterized by failure of a structural member subjected to high compressive stress. The criterion derived by Euler for determining the critical force in columns with no consideration for lateral forces is given by Equation \ref{EulerBuclking}. In the case of the CFRP tube the flexural rigidity (a product of the moment of inertia of the cross sectional area, $I$, and the modulus of elasticity, $E$) is particularly difficult to calculate due to the multiple orientations of the fibers on the tube and the characteristics of the epoxy. The CFRP tube's flexural rigidity is guaranteed by the manufacturer\footnote{CST composites} to be 43.61$\,$kN$\,$m$^{2}$. Considering the worst case scenario, which is the beam being fixed at both ends $\kappa=$0.50, the critical force for this beam is 325.45$\,$kN (73,162$\,$lbf) and corresponds to a safety factor of 3.79.
%
%
\begin{equation}\label{EulerBuclking}
    F = \frac{\pi^2(EI)}{(\kappa L)^2},
\end{equation}
where L is the length of the tube.

\begin{table}[htbp!]
\caption{Properties of the components of the \Spider\ spreader bar.}\label{mecha:spiderspreaderbartable}
\begin{center} 
\begin{tabular}{|l|c|l|c|c|c|c|}
\hline
    Component & Qty & Material & Mass & Design & Manufacturer & FOS  \\
     & & & (g) &  & & at 10g  \\ \hline
    Insert Joint& 2 & Al 7075-T6 &  2,301.93 & JDS & Quickparts & 1.28\\    
    Beam& 1 & CFRP & 2,268.29 & AA70430A & CST composites & 3.79 \\
    Shoulder bolts & 4& Steel & 735.44& 91259A313& McMaster Carr & \\
    \hline
\end{tabular} 
\end{center}
\end{table}

\section{Outer Frame}\label{mecha:outerframe}

\begin{figure}
\centerline{
\includegraphics[height=0.30\textheight]{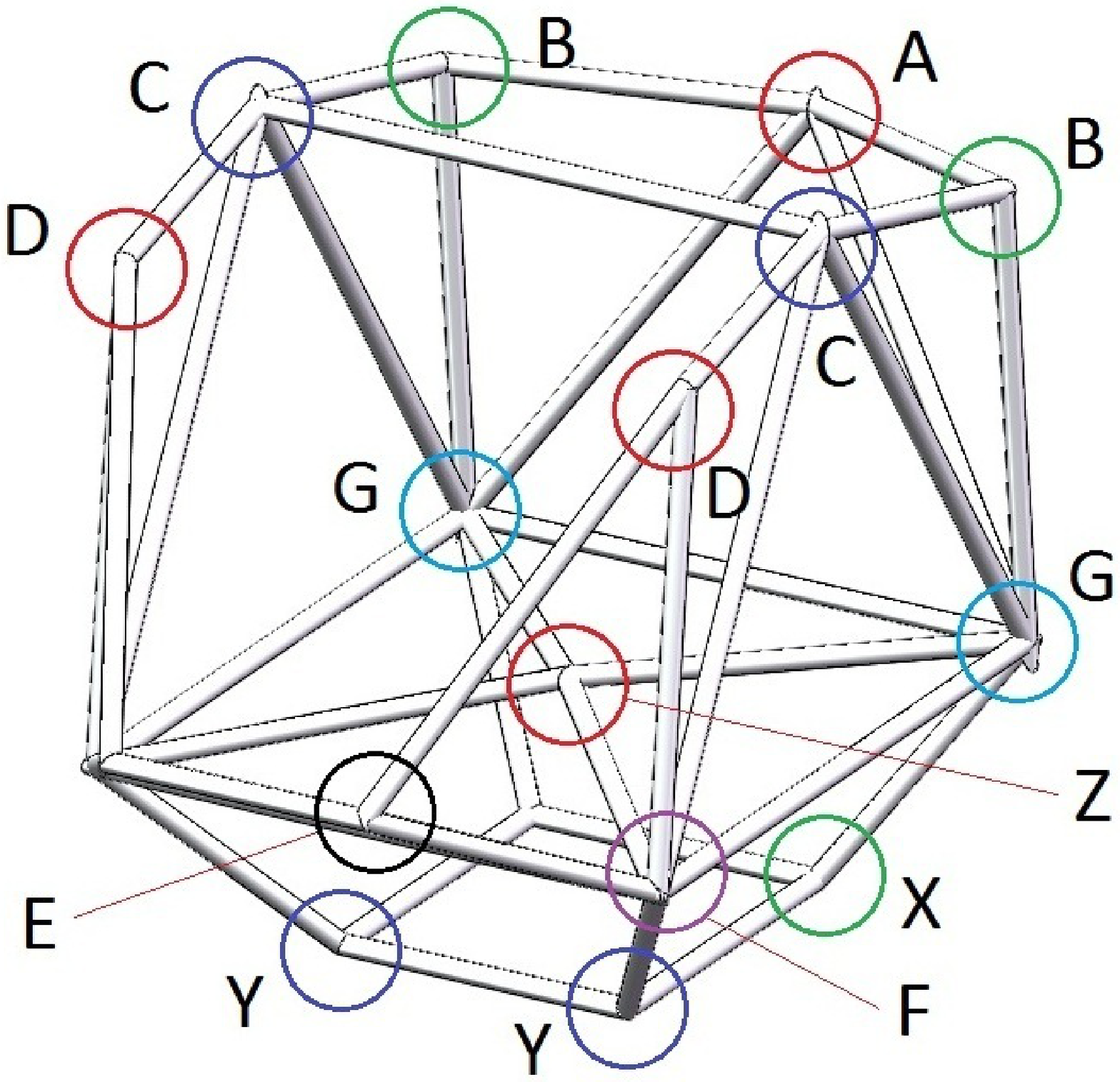}
\includegraphics[height=0.30\textheight]{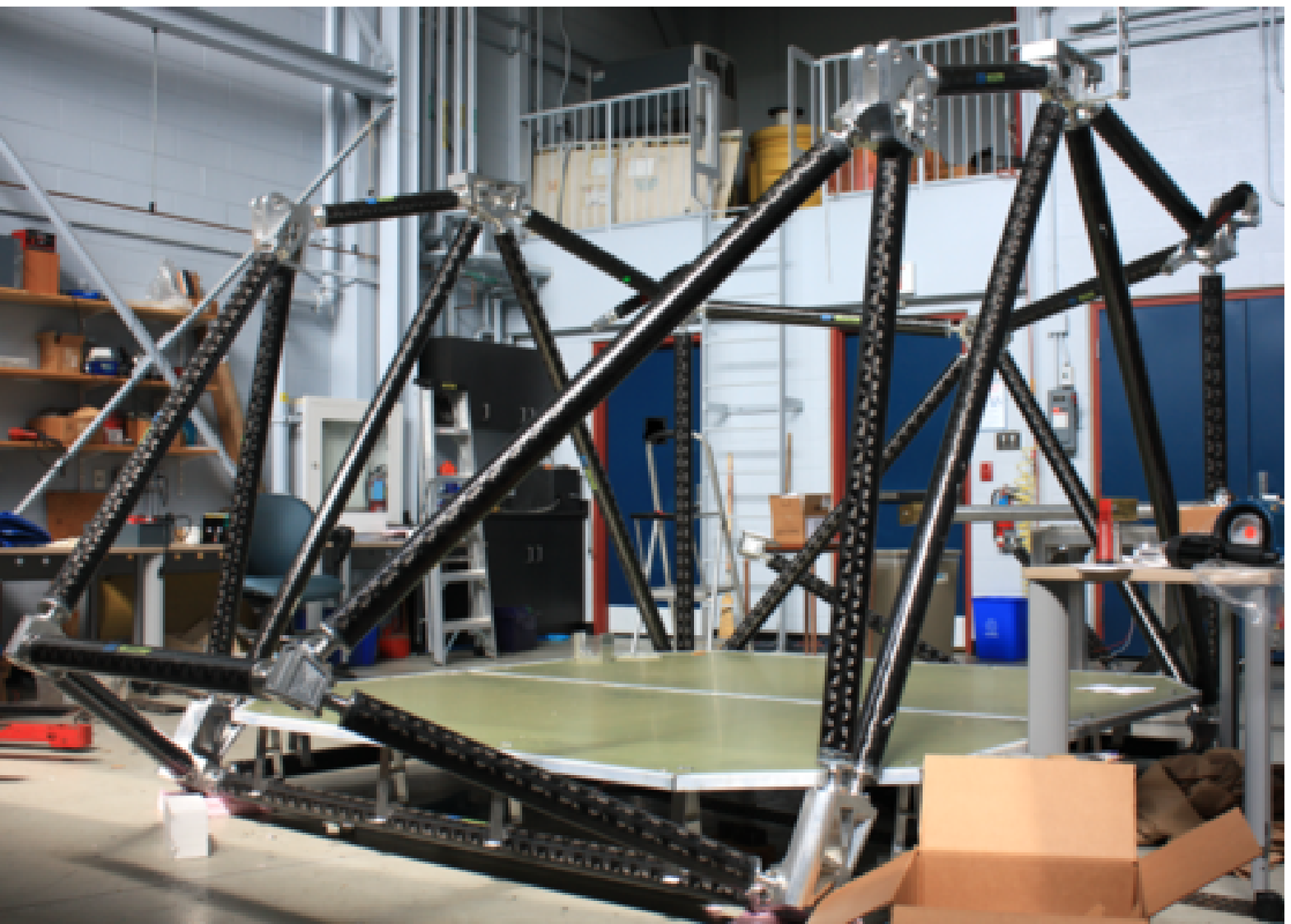}}
\caption{The \Spider\ gondola frame. Left: schematic showing the labeling of the multitube joints. Right: photograph of the \Spider\ gondola without the cryostat.}\label{OuterFrameJointFigure}
\end{figure}

The \Spider\ outer frame, shown in Figure \ref{OuterFrameJointFigure}, is made up of two kinds of components: CFRP tubes with aluminum inserts on their ends and multitube aluminum joints. The fastening of the inserts and the CFRP tube is made with an epoxy adhesive. The inserts are bolted into the faces of the aluminum joints that constrain the orientation angle of each tube. The final mass of the \Spider\ outer frame is 193.76$\,$kg (427.17$\,$lbs), accounting for only 9.4\% of the total mass of the experiment. The conceptual design of the \Spider\ outer frame is inspired by truss structures made with composite materials and it is largely influenced by the design of carbon fiber bicycles. The original evaluation of the CFRP tubing design\cite{martin_thesis2008} pointed out the difficulty of fastening multiple composite tubes and included several structures constructed with aluminum beams, such as the SIP cage and the reaction wheel support. The final version of the \Spider\ outer frame is entirely made of CFRP tubes held together with monolithic, custom-made multitube aluminum joints and a common aluminum insert glued to each end of the CFRP tubes.


The geometry of the outer frame is set by the scientific requirements of the experiment. The outer frame encloses the lower portion of the cryostat allowing its rotation. It also provides two square frames: one for mounting the deck with the attitude control electronics and one for supporting the SIP.

\subsection{Material Selection and Construction Technique}\label{mecha:technique}

\subsubsection{Carbon-Fiber-Reinforced Polymer (CFRP) Tubing}

CFRP is a very strong and light-weight material. Although it can be relatively expensive compared to aluminum and glass fiber, its high strength-to-weight ratio and good rigidity make it the perfect candidate for a light-weight structure such as the \Spider\ outer frame.

CFRP tubes are constructed by pultrusion, a continuous process for manufacture of composite materials with constant cross-section. The reinforced fibers are pulled through a resin, followed by a separate preforming system, and then into a heated die where the resin undergoes polymerization. The resin used for the \Spider\ outer frame tubes is epoxy. CFRP tubes are not weldable and drilling causes damage to the winding of the fibers, therefore they require metallic inserts to join them \cite{adams_adhesives1997}.

The CFRP tubes of the \Spider\ frame are provided by CST Composites, an Australian company with experience in manufacturing high quality filament wound tubing for industrial and marine applications. The selected product is the carbon/epoxy tubing AA70430A with 70.4$\,$mm inner diameter (ID) and a 3$\,$mm wall thickness. The linear density of this product is 1.056$\,$kg$\,$m$^{-1}$. It has a flexural rigidity (IE) of 43.61$\,$kN$\,$m$^{2}$ and maximum tensile strength of over 1,896.05$\,$MPa (275,000$\,$psi).

The selection of this product was made using a parametric static beam mesh study in Solidworks\tm. The parachute shock and landing scenarios (\ref{DesignScenarios}) were iteratively run on the model while changing the diameter and wall thickness of the tube until an optimal compromise between strength and estimated mass was achieved. According to FEA, the largest axial load on the final design of the \Spider\ outer frame is a 117.2$\,$kN (26,350$\,$lbf), resulting from the vertical pull of the front suspension cables at chute shock. Comparison of this axial load with the mechanical test performed on prototypes of the \Spider\ gondola tubes to test the adhesive joint results in minimum FOS of 1.72 at room temperature and 1.37 when pulled at lower temperatures in a bath of dry ice.

\subsubsection{Adhesive Joint}

The adhesive fastener selected for the \Spider\ outer frame is the 3M\tm Scotch-Weld\tm Epoxy Adhesive 2216 B/A Gray. This is a flexible, two-part, room-temperature-curing epoxy with high peel and shear strength. This product is recommended for bonding metals and plastics with good retention of strength after environmental aging and resistance to extreme shock, vibration and flexing. The Scotch-Weld\tm 2216 is widely used in aircraft and aerospace applications and it meets the DOD-A-82720 Military Specification for \emph{adhesive, modified-epoxies, flexible, and two-parts}.

The main difficulty in the construction of bonded joints is the assembly process. One of the main concerns in the case of the \Spider\ CFRP tube was galvanic corrosion. This can occur between aluminum and CFRP due to their large contact potential \cite{adams_adhesives1997}. Selecting materials with lower contact potentials like titanium and CFRP can minimize this effect. However, the higher specific strength and corrosion preventing features of titanium do not justify the greater cost, which is 5 to 10 times higher than aluminum \cite{callister_materials2005}. Instead the \Spider\ assembly avoids galvanic corrosion by ensuring electrical insulation between the two materials. This can be achieved by ensuring that the adhesive layer completely covers each overlapping surface thus allowing the use of aluminum for the inserts. The use of fishing lines as spacers was tested and dismissed because of the additional complexity in the assembly process and poor results in pull tests.

The final prescription for the surface preparation of the adhesive joint between the CFRP tubes and the aluminum joints is:

\begin{enumerate}
  \item Hand-sand the bonding surfaces of the aluminum insert to remove the corrosion layer in the metal and impurities on the surface.
  \item Hand-sand the inside of the CFRP tube without breaking the outer layer of epoxy, which keeps the carbon fibers together.
  \item Cover the edge of the outer face of the CFRP tube with masking tape to facilitate the cleaning of the tube. Clean both surfaces using acetone.
  \item Activate the glue by combining the two elements of the formula. 18~g of glue is enough for one outer frame insert. 3~g of glue is enough for one sunshield insert.
  \item Remove the bubbles trapped in the glue by putting the mix in a vacuum desiccator for two minutes.
  \item Cover both surfaces distributing the glue in uniform thin layers.
  \item Bring together both pieces, slightly rotating the insert.
  \item Install the tube in a clocking jig (in the case of the outer frame tubes) or in the assembly (in the case of the sunshield tubes).
\end{enumerate}

The working time of the Scotch-Weld\tm 2216 is 90~minutes. However, the glue thickens after 60 minutes of activation and it becomes harder to work with it. It is recommended to work in small batches to facilitate the gluing process. The curing of the outer frame tubes is made keeping the tube in a vertical position to avoid contact between the aluminum and the CFRP before the glue solidifies. The glue cures after 8 to 12 hours depending on the temperature conditions.

Because it is difficult to predict the strength and durability of adhesive joints, we experimentally pull-tested the assembled tubes. Quality assessment of each batch of glue is made by preparing a test assembly where the \Spider\ inserts are replaced by pull test inserts. The pull test of the first batch of tubes was performed at Sling-Choker MFG Ltd. where the adhesive joint was pulled to destruction when the load reached 201.1~kN (45,200~lbf) at room temperature and 161.2$\,$kN (36,230~lbf) when the tube was immersed in dry ice. This corresponds to a minimum FOS of 1.37 when compared to the largest axial load on the \Spider\ outer frame. The gluing of all the \Spider\ outer frame tubes was done between September and October 2009. Pull test of the control tubes is scheduled at different times before the deployment of \Spider\ to assess the adhesive joint after aging, stress on shipping, and exposure to sunlight.


\subsection{Inserts}

\begin{figure}
\centerline{\includegraphics[height=0.17\textheight]{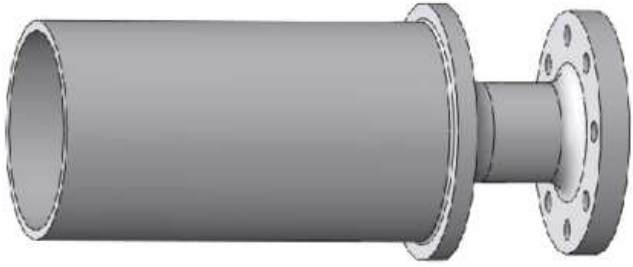}
\hspace{1cm}
\includegraphics[height=0.17\textheight]{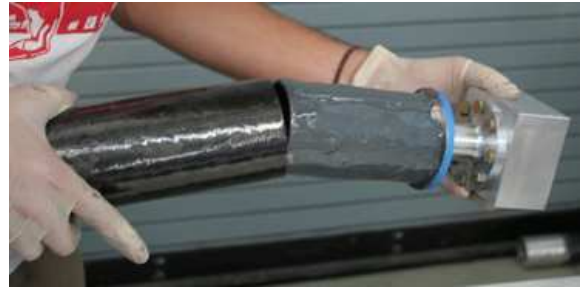}}
\caption{\Spider\ outer frame insert. Left: render of the insert. Right: preparation of the adhesive joint between the CFRP tube and the aluminum insert.}\label{Insert}
\end{figure}

The inserts, shown in Figure \ref{Insert}, are monolithic pieces of aluminum. Depending on their location on the outer frame, the inserts are made with either 6061-T6 or 7075-T6 aluminum. Both are tempered and precipitation hardened aluminum alloys. 6061-T6 is one of the most commonly used alloys of aluminum and it is often used in structure components of balloon-borne experiments. 7075-T6 has a yield strength comparable to many steels but it is more difficult to machine and considerably more expensive than 6061-T6. The selection of each type of aluminum depends on the particular force in each joint.

The inserts have two main sections: an adhesive joint section and a bolt fastening section. The adhesive joint is made between the cylindrical section of the insert and the CFRP tube. A fillet inside of the cavity reinforces the cylindrical section. The neck of the piece is machined to permit the location of the mounting bolts. All the inserts have a 3.175$\,$mm diameter canal that connects the interior of the cylindrical section with the mounting face and allows the air to circulate from the inside of the glued tube into the exterior. The FEA analysis of the inserts shows a FOS of 1.37 and 1.43 for the 6061-T6 and 7075-T6 aluminum inserts. The flange of the insert has been pull tested to destruction at 183.4$\,$kN (41,230$\,$lbf), which corresponds to a FOS of 1.79 with respect to the greatest axial stress during chute shock.

\subsection{Joints}

The CFRP are fastened in multi-tube joints in the nodes of the structure. The geometry of each multi-tube joint is determined by the orientation of the CFRP tubes set during the frame design. The joints are polyhedrons with faces perpendicular to the axes of the tubes. Each face fits the base of the insert and the eight-bolt pattern, thus the dimensions of the insert flange ultimately determine the overall size of the joint.

The diversity of angles in the outer frame required customized design for at least half of the tube connecting joints. In total, 13 different joints were designed for the outer frame, some of them with a mirrored version located on the other side of the gondola. The locations and names of the joints are shown in Figure \ref{OuterFrameJointFigure}. There are three joints connected to the suspension cables (A and D), two joints supporting the block bearings where the cryostat sits (C), four joints supporting the flywheel (Z), two joints supporting the elevation drive (B), four joints forming the frame for the deck and providing attachment points for the cart and the sunshields (F and G), and four joints forming the frame where the SIP sits.

All the aluminum joints include light-weighting pockets determined by FEAs. Each pocket reduces stress concentration in the joint while minimizing its mass. The final geometry of each pocket aims to simplify the machining process. However, the tubes orientations determine result in non-trivial joint geometries and their machining required a six-axis Computer Numerical Control (CNC) mill.

\subsection{Floor}

The floor of the \Spider\ gondola is the deck which supports the Attitude Control System (ACS) rack, the flight computers, the star cameras, the batteries and other electronics \cite{benton2014,gandilo2014}. The floor is made with two Teklam Corp. A510C 48~in$\times$96$\,$in and 1.0$\,$inch thick aluminum/aluminum honeycomb panels. These panels are manufactured with Teklam process specification TPS-A-A-500, which guarantees their ``\emph{use for primary and secondary aerospace applications}''. The density of the honeycomb panel is 4.54$\,$kg$\,$m$^{-2}$ (0.93~lb~ft$^{-2}$), which corresponds to a total mass of approximately 13.5$\,$kg (29.7$\,$lb) per section.

\begin{figure}
\begin{center}
    \includegraphics[height=0.25\textheight]{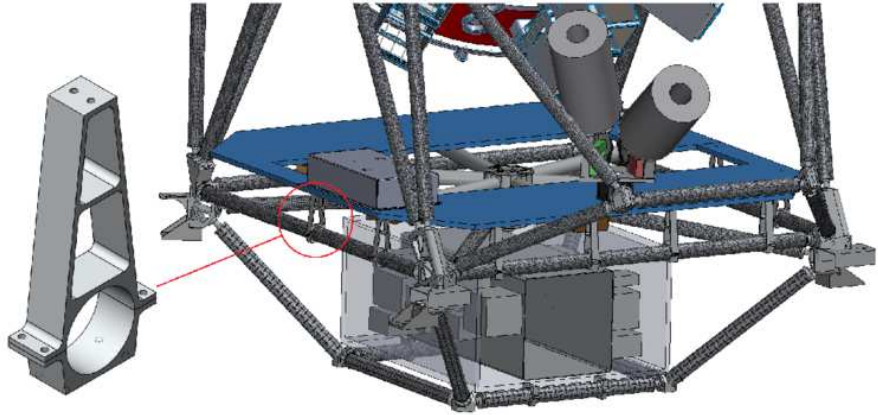}
    \caption{Rendering of \Spider's aluminum honeycomb deck and aluminum supports.}\label{mecha:spiderfloor}
\end{center}
\end{figure}

Each panel rests on eight aluminum supports. The lower ends of the aluminum supports fit around the CFRP tubes and are secured using a complementary bracket as shown in Figure \ref{mecha:spiderfloor}. The bottom bracket has a hole pattern to accommodate the line of sight transmitters.

\section{Sunshields}\label{mecha:sunshields}

The \Spider\ sunshield frame is mounted to the base of the outer frame using four aluminum supports. The frame is covered with two layers of aluminized mylar to protect the experiment from exposure to direct sunlight\cite{soler2014therm}. The \Spider\ sunshield frame is built using a combination of CFRP tubing and aluminum joints in a simplified version of the design used for the outer frame. This design was originally developed for the balloon-borne telescope BLASTPol and tested during the Antarctic flights of that experiment in 2010 and 2012\cite{fissel2010,galitzki2014}. 

\begin{figure}
\centerline{\includegraphics[height=0.28\textheight]{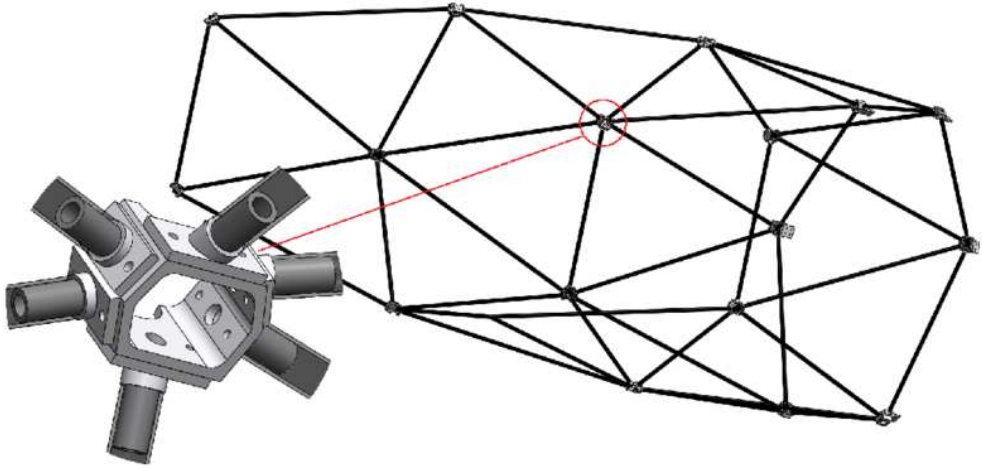}}
\caption[BLASTPol Baffle]{BLASTPol baffle assembly. The detail shows the multi-tube coupling to the aluminum hub.}\label{BLASTpolBaffle}
\end{figure}

The observation of the BLASTPol astronomical targets required shielding the telescope from direct sunlight at 40\degr\ from the sun position in azimuth. This was achieved with the construction of a baffle located around the 1.8$\,$m-diameter primary mirror and extending 7$\,$m in the bore-sight direction. The baffle frame is a truncated cylindrical structure with a hexagonal base made of CFRP tubes with aluminum inserts connected to aluminum hubs as shown in Figure \ref{BLASTpolBaffle}.

The CFRP tubes selected for thie BLASTPol baffle were the CST composites AA20020A: a 20.0$\,$mm ID and 2.0$\,$mm wall thickness CFRP tube. The selection of this product was made following a series of beam mesh simulations as described in Section \ref{mecha:outerframe}. Although the design constraints and loads are not as demanding as for the outer frame, the location of the baffle structure directly around the observation field of the telescope demanded special attention to the rigidity and structural integrity after launch. The structure required aggressive light-weighting to avoid increasing the moment of inertia of the cryostat over the capacity of the elevation drive. Additionally, the resonant frequencies of the structure had to be kept over $15\,$Hz to avoid resonance with the natural frequencies of the experiment.

The structure of the BLASTPol baffle starts with a hexagonal ring whose dimensions are determined by the clearance space around the primary mirror. Three more hexagonal rings are set at equal distances between the first ring and the minimum distance necessary to keep the sunlight off the primary and secondary mirrors at 40\degr\ in azimuth from the position of the sun. Each hexagonal ring is offset by 30\degr\ from the orientation of the previous ring to produce a triangular truss structure between each ring. 

The joints connecting the tubes are aluminum octahedrons with two parallel faces (top and bottom) and six faces perpendicular to the base forming a hexagon. They are centered on the node where the axes of the tubes converge, making the faces of the hexagon perpendicular to the tube axis. The inserts are made with aluminum plates, which mount to the hexagonal faces of the hub with two 1/4-20 bolts. The plates are welded to an aluminum cylinder oriented in the direction of the axis of the tube. This cylinder serves as the coupling surface for the adhesive joint with the CFRP tube as shown in the zoomed-in detail of Figure \ref{BLASTpolBaffle}.

The large area covered by the mylar makes the wind force an important aspect of the design of the sunshield frame. The magnitude of the wind force can be calculated from:
\begin{equation}\label{airforce}
    F_{air} = \frac{1}{2}\rho A C v^{2}.
\end{equation}
Where $A$ is the effective area facing the wind, $v$ is the velocity of the wind, and $C$ is the drag coefficient, which depends on the geometry of the shields\cite{batchelor2000introduction}. In order to validate the model, a worst-case scenario was considered where $A$ corresponds to the largest effective area of the shields and $C = 1$ corresponding to a flat surface perpendicular to the airflow. The BLASTPol structure has a FOS above 2.0 under the load of 20$\,$knot (10.3~m$\cdot$s$^{-1}$) wind.

The \Spider\ sunshields follow the same construction technique and design concept as the BLASTPol baffle: the structure is a cylinder with a hexagonal base composed of CFRP tubes fastened by aluminum inserts and joints. The main modification with respect to the design of the BLASTPol baffle is the diameter of the CFRP tubing, which has been increased to support larger loads. The chosen product is the CST Composites AA32018, a 32.0$\,$mm ID and 1.8$\,$mm wall thickness CFRP tube.

\begin{figure}
\begin{center}
    \includegraphics[height=0.35\textheight]{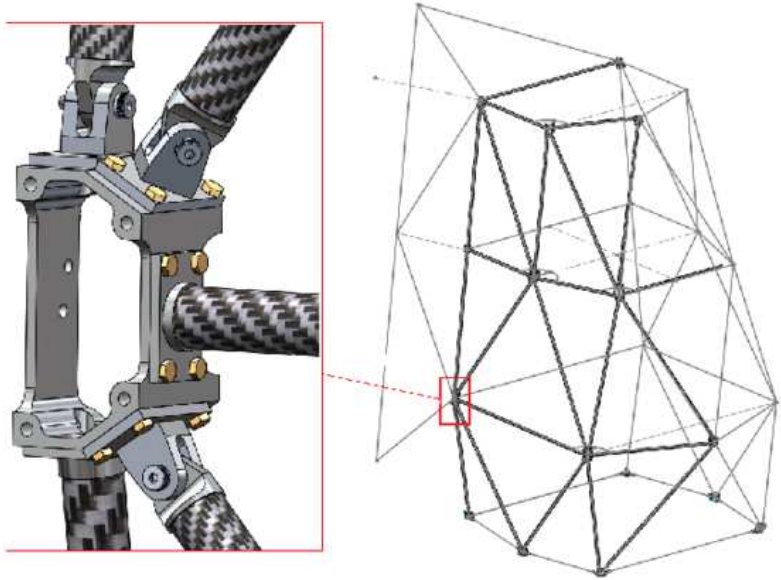}
\caption{\Spider\ sunshield assembly. The detail shows the multitube coupling at the aluminum hub.}\label{SpiderSunshields}
\end{center}
\end{figure}

The \Spider\ sunshields also implement a new feature in the construction technique: the combination of custom-angle fixed aluminum inserts with universal-joint-type insert, which can be oriented in two different angles as shown in Figure \ref{SpiderSunshields}. A universal-joint-type insert follows the tube incidence angle in two directions: one defined by the tube insert and yoke block hinge; the other is defined by the rotation of the yoke over the plate fastened to the aluminum hub. This versatile type of insert constitutes a general solution for the assembly of structures with moderate loads such as baffles and sunshields.

The gluing of this structure is made in place after \emph{dry} assembling the structure. The distance between the joints is constrained by cutting the tubes to length with $\pm$1$\,$mm tolerance. The assembly is temporarily held together using ropes around the tubes. In the absence of an extended flat surface to use as a reference for the assembly, the ropes constrain rotation of the hubs and ensure that the elements are held in the right configuration. The gluing is made triangle by triangle in several stages.

As for the BLASTPol baffle, one of the main considerations in the design of the \Spider\ sunshields is the lateral and vertical wind force. The sunshield frame is designed to hold the pivot during parachute shock and maintain structural integrity when covered with mylar and exposed to $20\,$knot ($10.3\,$m$\cdot$s$^{-1}$) wind. Its lowest resonance frequency is located over 15$\,$Hz, guaranteeing that the vibrational modes will not affect the pointing of the gondola.

\section{Conclusions}

A light-weight gondola is one of the characteristics that make \Spider\ unique as a suborbital experiment. Combining the use of composite materials and detailed FEA, the \Spider\ gondola pushes the limits fixed by the maximum gross lift of stratospheric balloons. Its initial Antarctic flight will measure or powerfully constrain the B-mode signal from primordial gravitational waves and provide a landmark in the design of balloon-borne telescopes.

\acknowledgments     

The \Spider\ collaboration gratefully acknowledges the support of NASA (award numbers NNX07AL64G and NNX12AE95G), the Lucille and David Packard Foundation, the Gordon and Betty Moore Foundation, the Natural Sciences and Engineering Research Council (NSERC), the Canadian Space Agency (CSA), and the Canada Foundation for Innovation. We thank the JPL Research and Technology Development Fund for advancing detector focal plane technology.  W.~C.~Jones acknowledges the support of the Alfred P. Sloan Foundation. A.~S.~Rahlin is partially supported through NASAs NESSF Program (12-ASTRO12R-004). J.~D. Soler acknowledges the support of the European Research Council under the European Union's Seventh Framework Programme FP7/2007-2013/ERC grant agreement number 267934. J.~D.~Soler thanks Taylor G. Martin and Marco P. Viero for their valuable comments on computer-aided design and carbon fiber gluing techniques. Logistical support for this project in Antarctica is provided by the U.S. National Science Foundation through the U.S. Antarctic Program. We would also like to thank the Columbia Scientific Balloon Facility (CSBF) staff for their continued outstanding work.

\bibliography{InstrumentationRefs,jdslib,CMBrefs,ISMrefs}{}   
\bibliographystyle{spiebib}   

\end{document}